\documentclass[twocolumn]{jetpl}
\usepackage[utf8]{inputenc}
\usepackage[T1,T2A]{fontenc}
\usepackage{cite}
\usepackage{physics}
\usepackage{mathtools}
\usepackage{xcolor}
\usepackage{hyperref}

\lat

\title{SiPM-based photodetectors for spectroscopic measurements in wide dynamic range}

\rtitle{SiPM-based photodetectors for spectroscopic measurements in wide dynamic range}

\sodtitle{SiPM-based photodetectors for spectroscopic measurements in wide dynamic range}

\author{
  S.\,V.\,Evdokimov$^{+}$, A.\,M.\,Gorin$^{+}$,
  V.\,I.\,Izucheev$^{+}$,
  Y.\,V.\,Kharlov$^{+\ddagger}$\thanks{Corresponding author: Yuri.Kharlov@ihep.ru}, 
  B.\,V.\,Polishchuk$^{+}$, V.\,I.\,Rykalin$^{+}$, 
  S.\,A.\,Sadovsky$^{+}$,
  A.\,A.\,Shangaraev$^{+}$
  N.\,E.\,Voronkov$^{+}$, 
  A.\,A.\,Zaitsev$^{\dagger}$
}

\rauthor{Evdokimov, Gorin, Izucheev, et al.}

\sodauthor{
  S.\,V.\,Evdokimov$^{+}$, A.\,M.\,Gorin$^{+}$,
  V.\,I.\,Izucheev$^{+}$, Y.\,V.\,Kharlov$^{+\ddagger}$\thanks{e-mail:Yuri.Kharlov@ihep.ru}, 
  B.\,V.\,Polishchuk$^{+}$, V.\,I.\,Rykalin$^{+}$, 
  C.\,A.\,Sadovsky$^{+}$,
  A.\,A.\,Shangaraev$^{+}$
  N.\,E.\,Voronkov$^{+}$, 
  A.\,A.\,Zaitsev$^{\dagger}$
}

\address{
  ~\\$^+$NRC ``Kurchatov Institute'' -- IHEP, Protvino,
  Moscow region, 142281, Russia
  \\
  ~\\$^\dagger$JINR, Dubna, Moscow region, 141980, Russia
  \\
  ~\\$^\ddagger$MIPT, Dolgoprudny, Moscow region, 141701, Russia
}

\dates{25 December 2025}{*}

\abstract {
  The results of measurements of
  the characteristics of photodetectors based on large-area SiPM
  matrices with a wide dynamic range are presented. The potential
  application of SiPM-based photodetectors for electromagnetic
  calorimetry in the energy range from hundreds of keV to tens of
  GeV is explored. The potential application of SiPM in
  gamma-spectroscopy is also discussed.
} 


\begin{document}

\maketitle

\section*{Introduction}

The measurement of weak light signals in the
visible optical range is a common problem in many areas of
physics. The light flux is converted into an electrical signal,
measured by photodetectors, and the magnitude of the current or charge
induced by the photodetector determines the magnitude of the incident
light flux. 

In experimental nuclear physics and high-energy physics,
ionizing radiation is detected using visible and near-ultraviolet
light signals, the intensity of which is proportional to the energy
loss of high-energy particles passing through the detector's active
medium.
In particular, in electromagnetic and hadronic calorimeters, particles
release almost all of their initial energy in the calorimeter
medium, and secondary charged particles, emitted from 
interaction of the primary particle with the material, excite the
active medium of the calorimeters, for example, scintillators or
Cherenkov radiators.
The energy released by secondary particles in these active media
determines the light flux emitted by these media, so the luminous flux
of the recorded light pulses or the charge induced by the
photodetector is uniquely related to the released energy, and hence to
the initial energy of the incident high-energy particle.
Another class of detectors that require recording and measuring the
magnitude of a light signal are scintillation counters of charged
particles.
When particles pass through scintillation counters, ionization losses
depend weakly on the particle velocity, but strongly on the particle
charge. With sufficient accuracy in measuring the current or charge
induced by the photodetector, scintillation detectors can determine
both the charge and energy of particles. When combined with magnetic
spectrometers that measure the momentum of a charged particle, they
can determine its mass, i.e., the particle type.
These tasks impose requirements on the basic characteristics of
photodetectors that measure the charge from short and weak light
pulses emitted from scintillation detectors.
These main characteristics are the efficiency of photon detection,
the resolution of photodetectors for the measured light flux, the
stability of the amplitude measurement over time, and the degree of
linearity of the measured amplitude of the light pulse from the
initial energy released by the particle in the active medium of the
detector.
The typical range of induced charges from light pulses emitted by
calorimeters and scintillation counters used in high-energy and
nuclear physics ranges from tens to hundreds of thousands of
photoelectrons. The duration of such light pulses is quite short, on
the order of nanoseconds. 

Traditionally, electrical signals from weak light signals are measured
using vacuum photomultipliers. Advances in semiconductor technology
over the past two decades have led to the development of silicon
photodiodes operating in the Geiger mode, known as silicon
photomultipliers (SiPMs) or multi-pixel photon counters (MPPCs). The
term SiPMs will be used throughout this paper.  SiPMs have already
proven themselves in scintillation
detectors~\cite{Brekhovskikh:2024qqj} in counting mode, where the
requirements for light flux measurement accuracy and dynamic range are
relatively soft. SiPMs are also used in a number of applications for
recording single-photon light signals~\cite{Balygin2017.en}.

To measure light signals over a wide amplitude range, SiPMs with a
large number of pixels are required, and as the light flux density
increases, the pixel size must be reduced to minimize the probability
of multiple photons hitting a single pixel.
Modern experiments in high-energy physics consider SiPM as a direction
for upgrading photodetectors for hadronic~\cite{CMS:2017jpq} and
electromagnetic calorimetry~\cite{Balygin:2018bgj}.
The required range of induced charges from the active detecting
element of the calorimeter, measured in photoelectrons (p.e.), can be
estimated using the example of the PbWO$_4$ scintillating crystal
proposed for the electromagnetic calorimeter of the PANDA
experiment~\cite{Borisevich:2005wb}.
The light yield of a single full-size crystal of this calorimeter is
at least 20 p.e./MeV. To measure photon energies in the range from
10~MeV to 10~GeV, it is necessary to register signals in the range
from 200 to $2\times10^5$~p.e.
Since the signal amplitude from the SiPM is proportional to the number
of pixels activated by incident light, the SiPM requires a pixel count
of several hundred thousand. More precise requirements for the SiPM
size, the size and number of pixels, are determined by the specific
calorimeter features such as the transverse dimensions of the detecting
element, the light yield, the required energy resolution, and the
energy range defined by the physical objectives of a given experiment.

There are SiPMs available on the market from Hamamatsu and Capital
Photonics Technology (Tianjin) Co., Ltd., that can be used for
spectrometric measurements. Table~\ref{tab:SiPM} presents several
SiPMs with an area of $6.0 \times 6.0~\mbox{mm}^2$ and a pixel size of
10 to $20~\mu$m, which are considered promising photodetectors for
amplitude measurements of light pulses over a wide dynamic range. In
this paper, we investigate the characteristics of one of these SiPMs,
namely the EQR15 11-6060D-S from NDL, with a pixel size of $15~\mu$m and a
pixel count of over 170,000.
\setcounter{table}{0}
\begin{table*}[ht]
  \centerline{
  \begin{tabular}{l|c|c|c|c} \hline
    Peremeter & Hamamatsu      & Hamamatsu      & NDL EQR15   & ZJGD EQR20 \\
             &  S14160-6010PS &  S14160-60105S & 11-6060D-S  & 11-6060D-S \\
             & \cite{Hamamatsu} & \cite{Hamamatsu} & \cite{NDL} & \cite{ZJGD} \\
    \hline
    size, mm$^2$ & $6.0 \times 6.0$ & $6.0 \times 6.0$ & $6.24
    \times 6.24$ & $6.24 \times 6.24$ \\ \hline
    pixel size, $\mu$m & 10 & 15 & 15 & 20 \\ \hline
    number of pixels & 359011 & 159565 & 173038 & 97344 \\ \hline
    Break voltage $V_{\rm br}$, В & $38 \pm 3$ & $38 \pm 3$ & 30 & $27.2 \pm 1$ \\ \hline
    Working voltage, V & $V_{\rm br}+5$ & $V_{\rm br}+5$ & $V_{\rm br}+8$ &  $V_{\rm br}+5$ \\ \hline
    photon detection efficiency, \% & 18 & 32 & 45 & 47.8 \\ \hline
    gain & $1.8 \times 10^5$ & $3.6 \times 10^5$ & $4.0 \times 10^5$ & $8.0 \times 10^5$ \\ \hline
    dark count rate, $\mbox{s}^{-1}$ & $(3-10) \times 10^6$ & $(3-10) \times 10^6$ &
    $10^7$ & $(6-16) \times 10^6$ \\ \hline
    Wave length at  max. efficiency, nm & 460 & 460 & 420 & 420 \\ \hline
    temperature coefficient, mV/$^\circ$C & 34 & 34 & 28 & 24.8 \\ \hline
    capacity, pF & 2200 & 2200 & 218 & 397 \\ \hline
  \end{tabular}
  }
  \caption{Tab.1. Comparison table of spectrometric SiPM available on
    the market.} 
  \label{tab:SiPM}
\end{table*}
%

\section{Photodetectors based on SiPM matrices}

The requirements for photodetectors for recording light pulses from
electromagnetic calorimeter cells are determined by the design and
manufacturing technology of the calorimeter, the light yield of the
calorimeter radiators, the energy range of the primary photons, and
the energy resolution determined by the physical objectives of the
experiment.
If one considers homogeneous calorimeters, for example, made of lead
glass or inorganic scintillators, then the signal is recorded through
the end cap of the calorimeter cells.
The transverse cell size is, in turn, determined by the Moli\`ere radius
of the calorimeter's radiator material. Typical electromagnetic
calorimeters, common in high-energy physics, have transverse cell
sizes ranging from $22-26$~mm, as in calorimeters based on lead
tungstate crystals, to $40-85$~mm, as in calorimeters with lead glass
radiators.
The category of homogeneous detectors also includes MeV-range
gamma-ray spectrometers based on scintillating crystals with high
light yields, such as NaI(Tl), CsI(Tl), GAGG, and others. The active
area of the photodetector,
mounted at the end cap of the calorimeter or
$\gamma$-spectrometer cell, directly determines the light collection
efficiency and, consequently, the energy resolution of the entire
detector. Therefore, to increase light collection from detector
radiators, it is natural to use SiPM matrices.

\begin{figure}[htb]
  \centering
    \includegraphics[width=0.9\linewidth, bb=140 120 390 290]{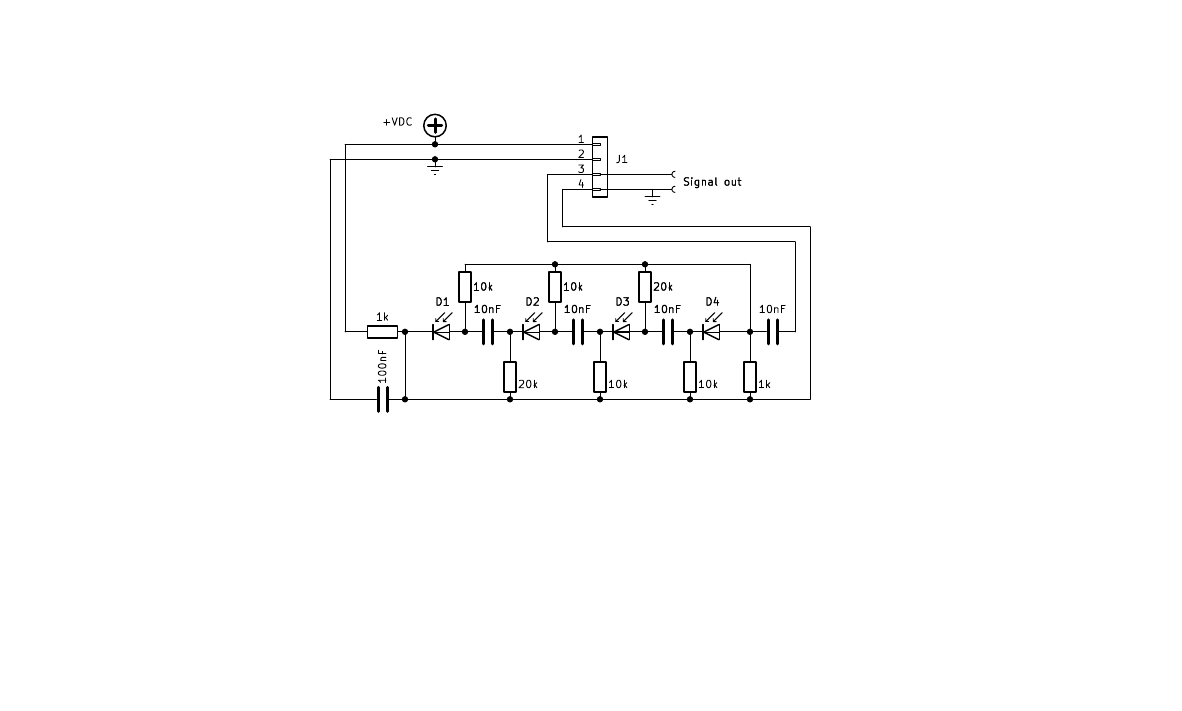}
  \caption{Fig. 1. Schematic diagram of combining four SiPMs into a photodetector matrix.}
  \label{fig:SiPM4_schematics}
\end{figure}

Several schemes for combining multiple SiPMs into a single
photodetector are discussed in the literature. The so-called hybrid
scheme for connecting multiple SiPMs is considered optimal, in which
the bias voltage is applied in parallel to all SiPMs, and the signals
from individual SiPMs are collected sequentially through decoupling
capacitors~\cite{Bonesini:2023hzg}.
The schematic for combining four SiPMs into a single photodetector
array, which was studied in this work, is shown in
Fig. ~\ref{fig:SiPM4_schematics}. When combining commercial SiPMs from
NDL, presented in Table~\ref{tab:SiPM}, such a photodetector will
provide light collection from an area of
$12.5\times12.5~\mbox{mm}^2$. Given the high quantum efficiency of
SiPMs, this already allows one to consider a $2\times2$ SiPM array as
an alternative to vacuum photomultipliers. 
%

\section{Selecting the operating mode of SiPM-based photodetectors}

\begin{figure}[htb]
    \centering
    \includegraphics[width=\linewidth]{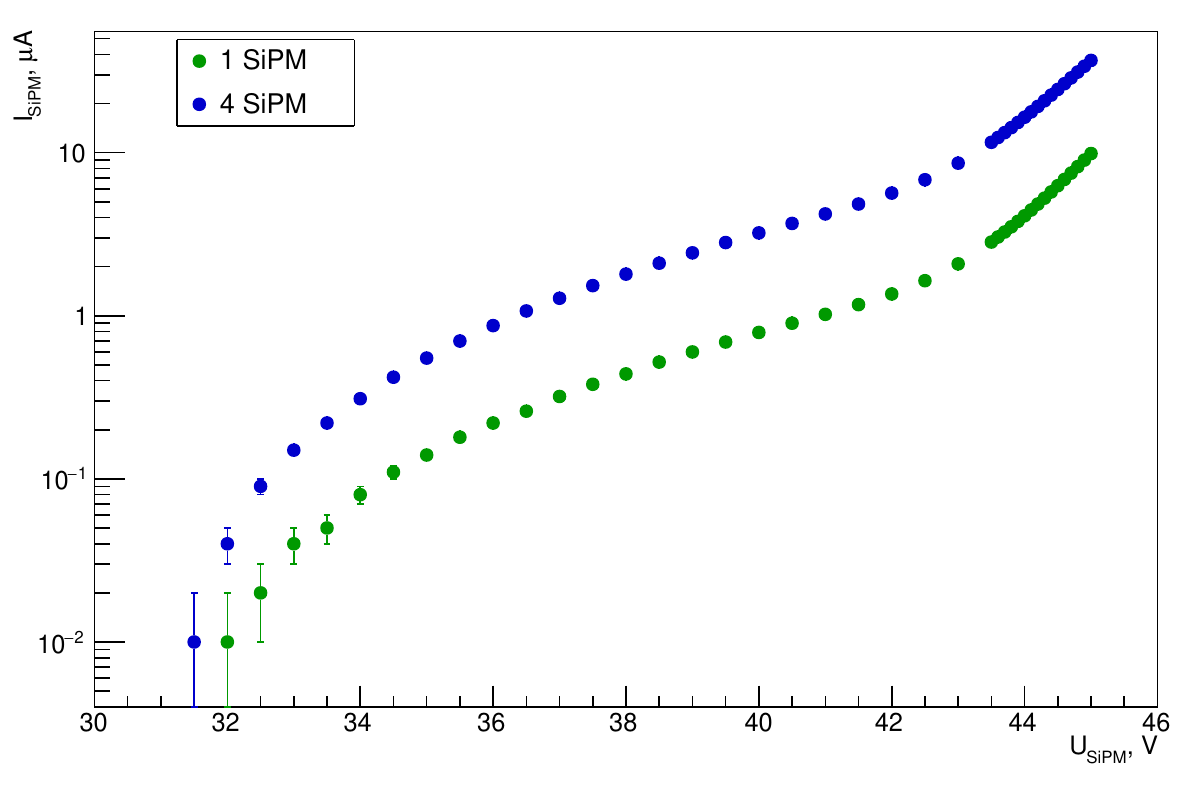}
    \caption{Fig. 2. current-voltage characteristic for photodetectors
      of 1 and 4 SiPM.
    }
    \label{fig:VACh}
\end{figure}

The vendor of the SiPM EQR15 11-6060D-S specifies a recommended
operating voltage of $V_{\rm br}+8$~V~\cite{NDL}. To determine the
optimal bias voltage for spectrometric measurement applications, it is
necessary to measure the photodetector's current-voltage
characteristic, as well as the dependence of its gain and amplitude
resolution on bias voltage.
Fig. ~\ref{fig:VACh} shows the measured current-voltage characteristic
for two photodetectors, one consisting of a single SiPM and the other one
consisting of four SiPMs connected in a hybrid circuit. For both
photodetectors, the breakdown voltage at which current begins to flow
through the SiPMs is approximately $31-32$~V. At voltages above 43~V,
the current-voltage curve changes the slope, and the current increases
sharply.

\begin{figure}
    \centering
    \includegraphics[width=0.49\linewidth]{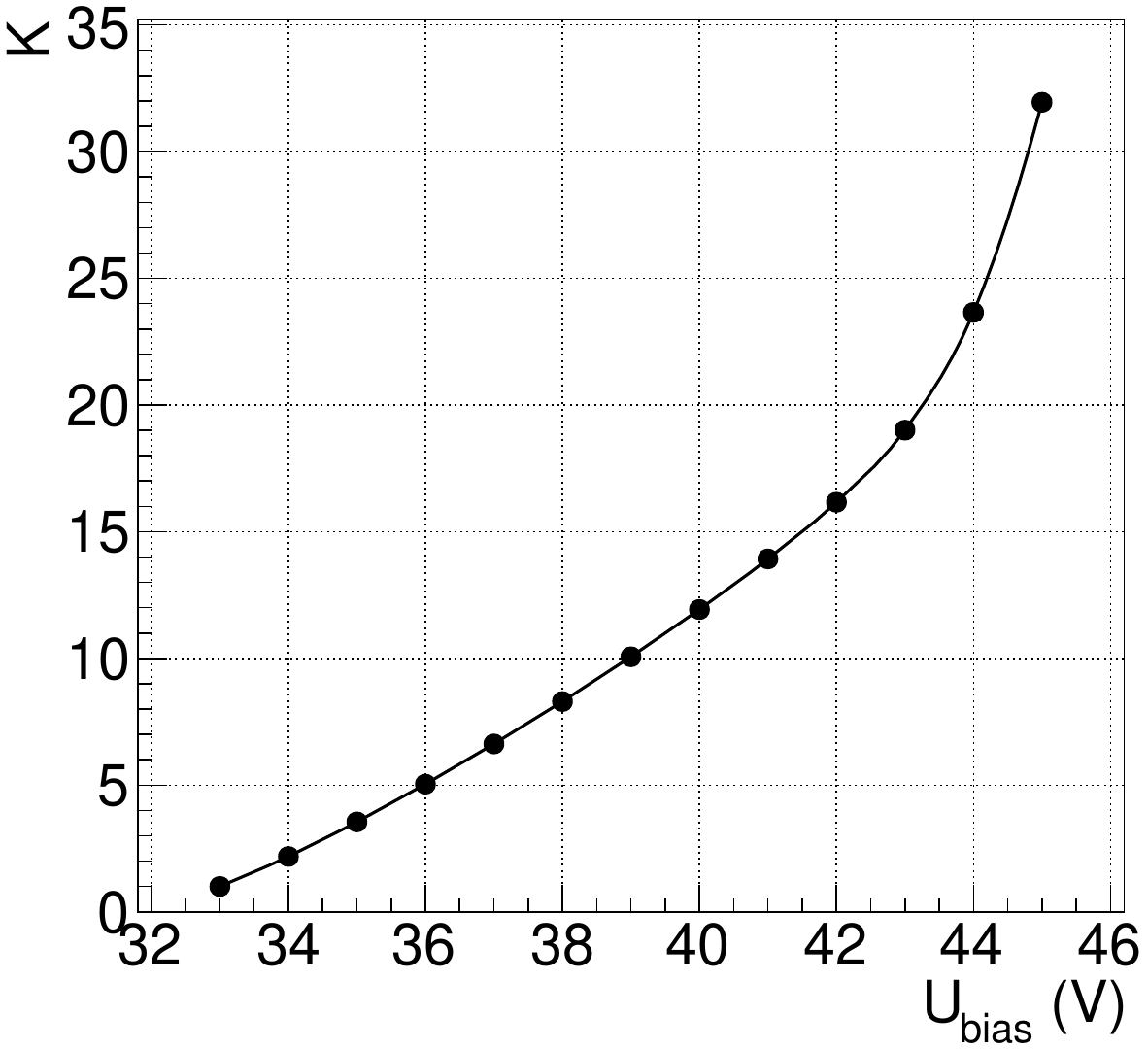}
    \hfill
    \includegraphics[width=0.49\linewidth]{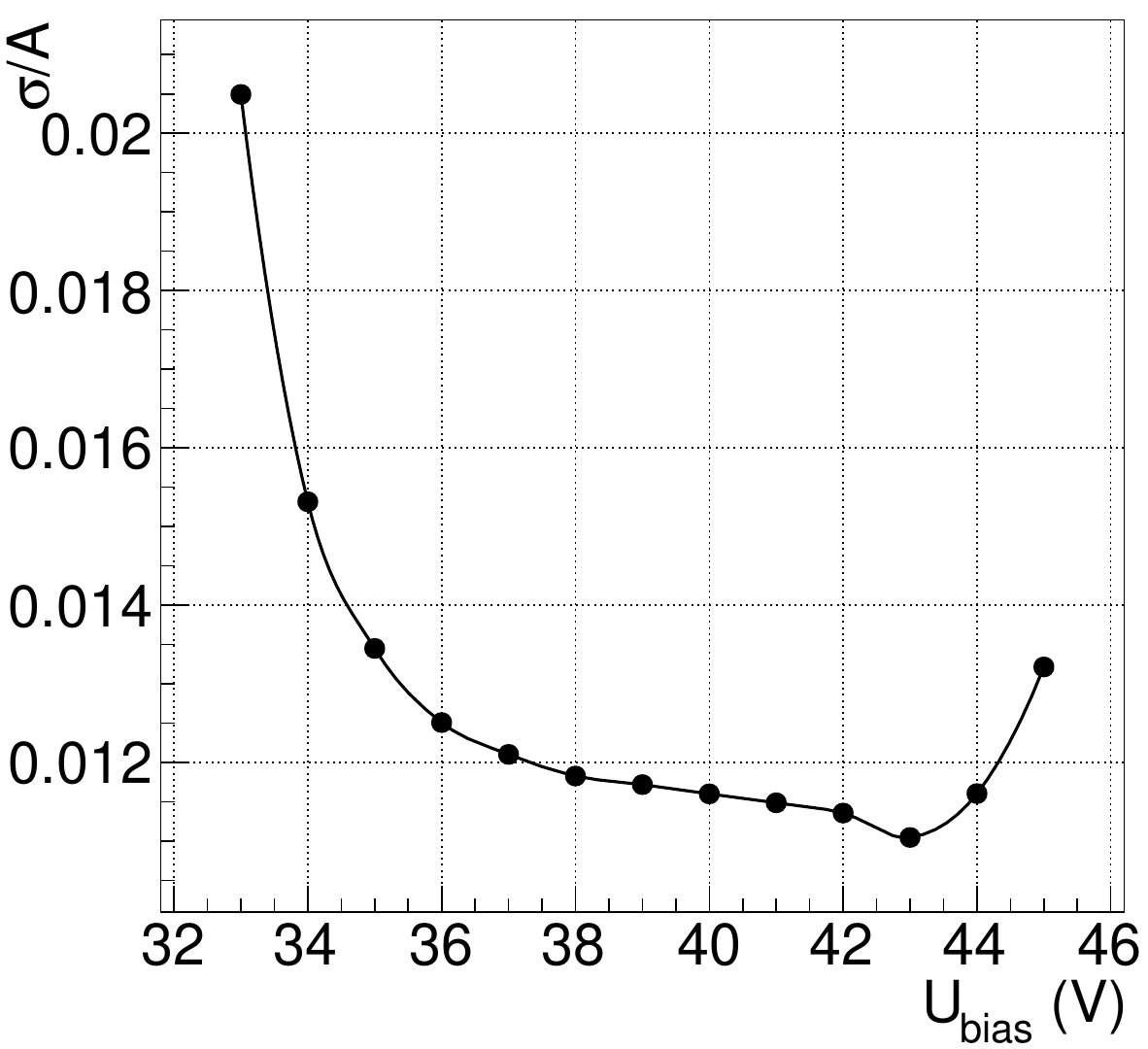}
    \caption{Fig. 3. Dependences of the relative gain (left) and
      relative amplitude resolution (right) of a SiPM-based
      photodetector on the bias voltage.
    }
    \label{fig:SiPM1_gain_resol}
\end{figure}

The average amplitude and its standard deviation were also measured as
functions of the applied bias voltage in the range from 33~V to
45~V. A LED pulse generator with feedback, ensuring a luminous flux 
stability of no worse than 0.3\%, was used as a pulsed light
source. The generator's pulse duration was $8-20$~ns, the emitted light
wavelength was 460~nm, the pulse rate was 1000 Hz, and the
amplitude was specified by a 12-bit digital-to-analog converter (DAC)
code.
The amplitude spectra of the measured signals from the photodetector
at each given voltage $U_{\rm bias}$ were fitted with a Gaussian
function, from which the average value $A$ and the standard deviation
$\sigma$ were determined. The relative gain $K$ was measured at a
fixed amplitude of the LED pulse generator and was defined as the
ratio of the average amplitude $A_U$ of the signal from the
photodetector at a given voltage $U_{\rm bias}$ to the amplitude
$A_{33}$ at a voltage $U_{\rm bias}=33$~V, $K=A_U/A_{33}$.
Fig. ~\ref{fig:SiPM1_gain_resol} shows the dependences of the relative
gain $K$ and the relative amplitude resolution $\sigma/A$ on voltage
$U_{\rm bias}$, which demonstrate a change in behavior at voltages
above 43~V. It is important to note that at voltages significantly
above 43~V, the dependence of the SiPM gain on voltage becomes
significantly nonlinear, and the resolution deteriorates. Thus, both
the current-voltage characteristic and the gain and amplitude
resolution indicate that the acceptable operating voltage for the SiPM
lies in the range from the breakdown voltage up to 43~V.

\section{Linearity of SiPM-based photodetector}

For spectrometric detectors, linearity, i.e. the proportional
dependence of the measured signal of the photodetector on the incident
light flux, is an important characteristic.
However, possible photodetector nonlinearity is not a significant
obstacle to its use in calorimeters in high-energy physics or in gamma
spectroscopy in nuclear physics, as the detectors themselves may
exhibit minor nonlinearity caused by incomplete energy deposition of the
primary photon in the calorimeter radiator and the threshold of the
readout electronics. Therefore, photodetector nonlinearity, if
present, must be measured separately for subsequent calibration of the
measured signal relative to the incident photon energy.

To measure the dependence of the signal amplitude from a photodetector
consisting of four SiPMs on the incident light flux, the LED pulse
generator described in the previous section was used. The incident
light flux onto the photodetector was determined by the generator's
DAC code.
Since the LED generator itself can be nonlinear, an independent
linearity check was carried out on the signal amplitude generator code
from the FEU-84-3 photomultiplier, which was widely used previously in
electromagnetic calorimeters and has proven itself as a spectrometric
photodetector with a wide dynamic range, having a linear response up
to signal amplitudes of $\approx 1.3$~V on a 50~Ohm
load~\cite{Serpukhov-Brussels-AnnecyLAPP:1985jek}.  

\begin{figure}
    \centering
    \includegraphics[width=0.49\linewidth]{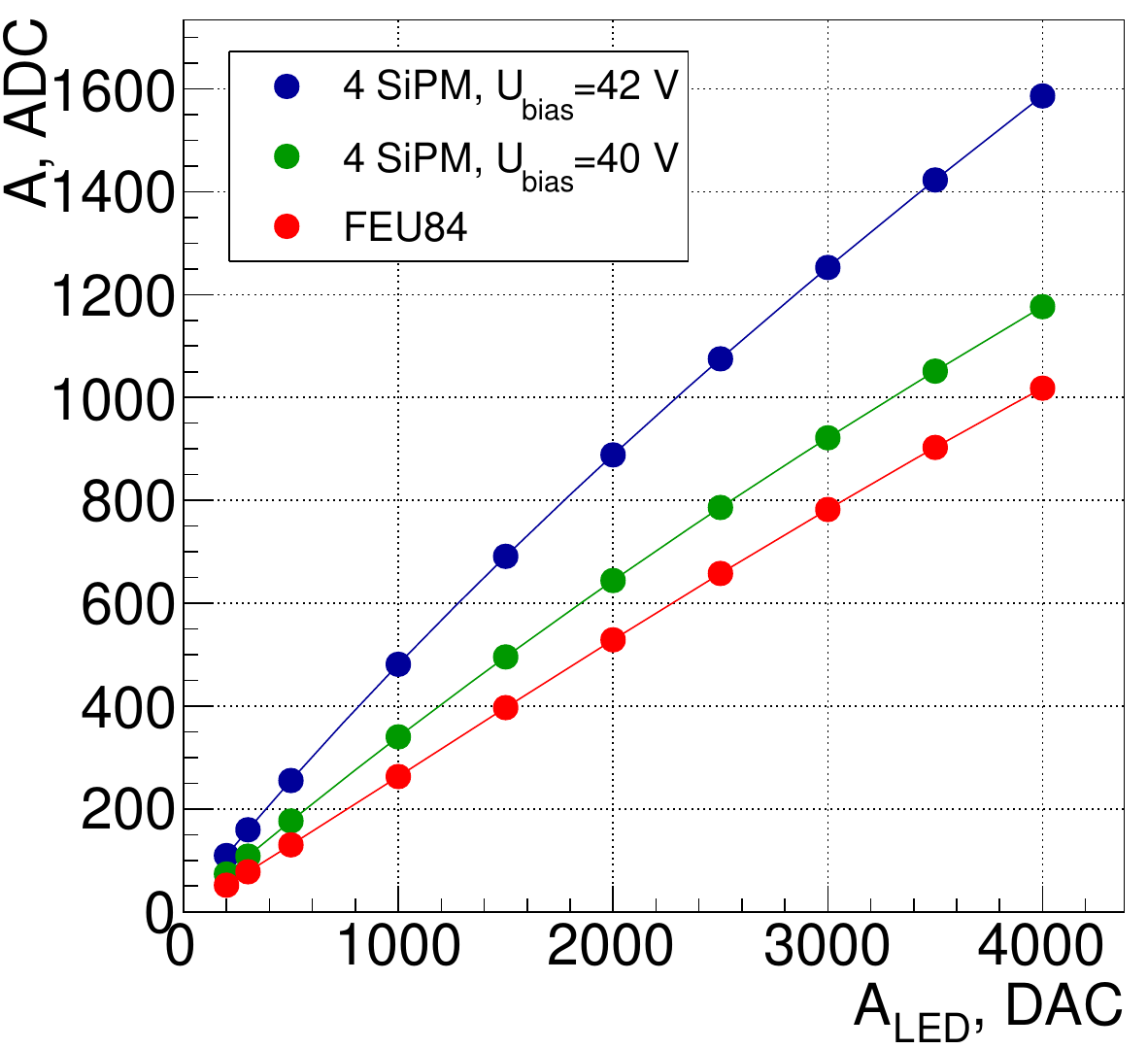}
    \hfill
    \includegraphics[width=0.49\linewidth]{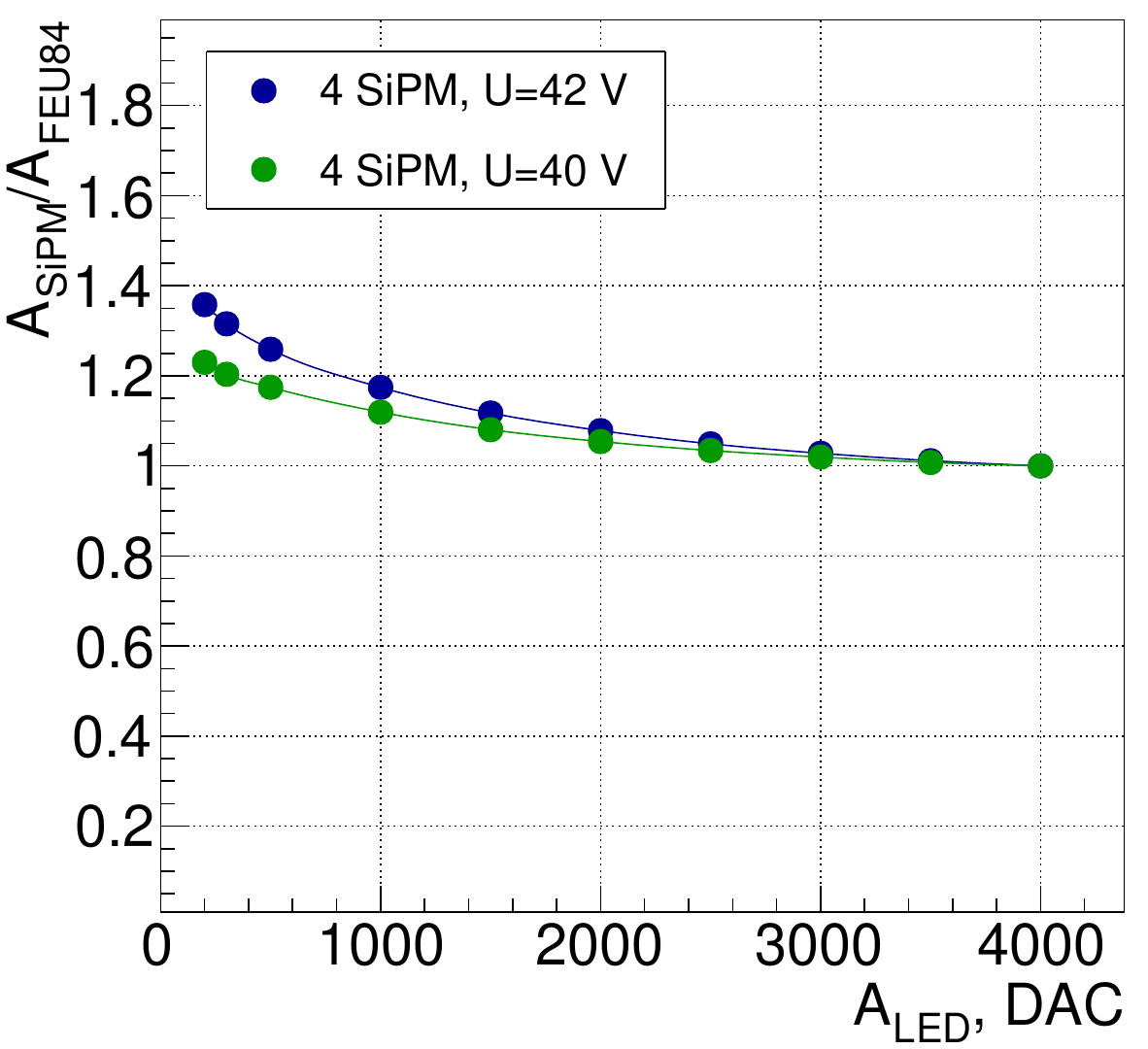}
    \caption{Fig. 4. Measured dependences of the signal amplitude from
      a photodetector based on 4 SiPM at voltages of 40 and 42~V and
      with FEU-84-3 (left) and the ratio of the amplitudes of 4 SiPM
      to FEU-84-3 (right) on the DAC code of the LED pulse generator.
    }
    \label{fig:SiPMvsFEU84}
\end{figure}

The left Fig.~\ref{fig:SiPMvsFEU84} shows the dependences of the
amplitudes of signals from a photodetector of four SiPMs at a bias
voltage of 42~V and from a photomultiplier FEU-84-3 at a voltage of
1700~V, on the DAC code of the LED generator $\rm{A}_{\rm LED}$.
The ratio of the amplitudes of the 4-SiPM photodetector at bias
voltages of 40 and 42~V to the FEU-84-3 amplitude, normalized to the
ratio at maximum luminous flux, is shown in the right
Fig.~\ref{fig:SiPMvsFEU84}. A slight nonlinearity of the response
of the 4-SiPM photodetector relative to the FEU-84-3 is observed at a
level of 20\% at $U_{\rm bias} = 40$~V and 25\% at $U_{\rm bias} =
42$~V in the range of luminous fluxes differing by more than 20
times. This relative nonlinearity of the 4-SiPM photodetector,
measured using a LED generator, can be used for amplitude correction.

\section{Amplitude resolution of SiPM-based photodetectors}

For precision spectrometric measurements, detector amplitude
resolution is one of the key characteristics by which various
detectors of this class are compared. One factor contributing to the
amplitude resolution of a spectrometric detector is the internal noise
of the measurement electronics, specifically the noise of the
photodetector, amplifier, and signal digitization electronics.
This noise contribution is generally independent of the signal
amplitude. When recording weak light signals, the stochastic
contribution to the detector's resolution, i.e., the contribution
determined by stochastic fluctuations of a finite number of
photoelectrons, can be dominating.
The accuracy of the light source's amplitude setting can also affect
the measured amplitude resolution of the photodetector under
study. The resolution of the SiPM-based photodetector was studied
using an LED generator, whose contribution to the measured resolution
is no more than 0.3\%. Therefore, the effect of generator instability
on the resolution is negligible. 

The amplitude of the measured signal is proportional to the number of
detected photons (photoelectrons), so the stochastic contribution to
resolution is proportional to the square root of the amplitude. With a
sufficiently broad light flux, the number of detected photons is
directly determined by the area of the photodetector.

\begin{figure}
  \centering
    \includegraphics[width=\linewidth]{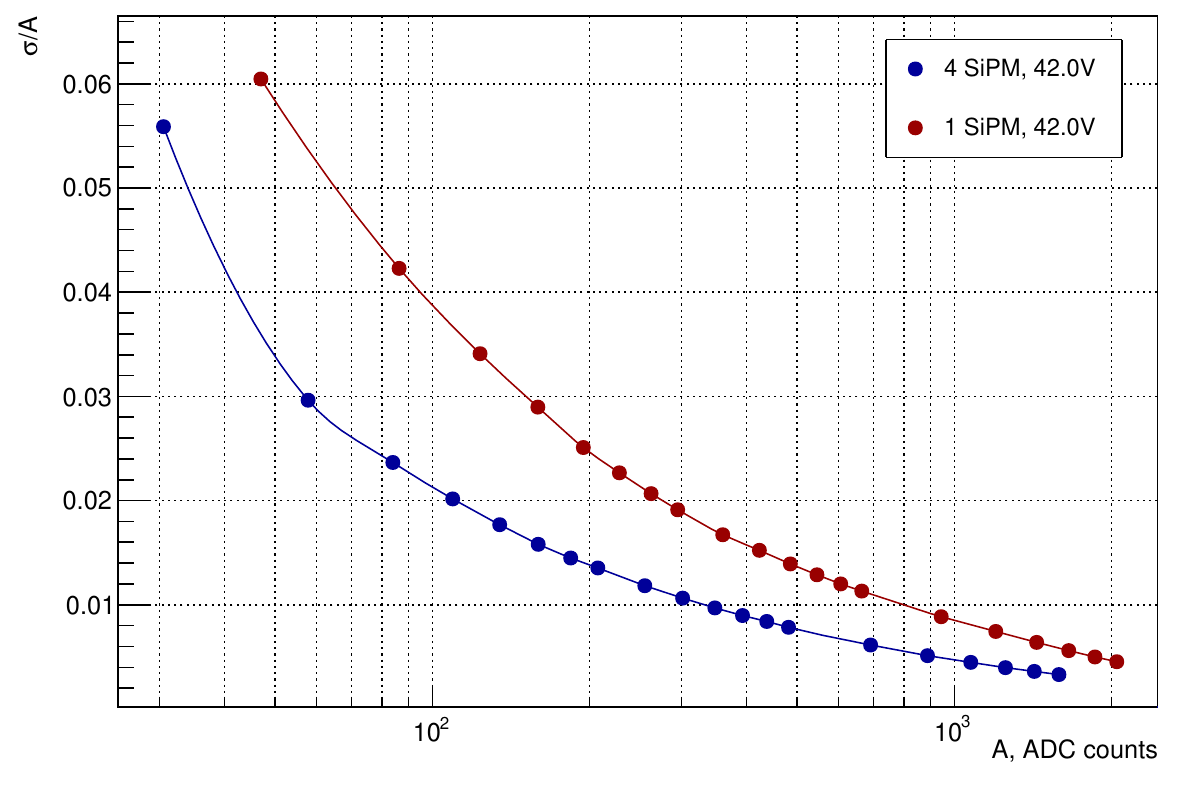}
    \caption{
      Fig. 5. Dependences of the relative amplitude resolution of
      photodetectors from one and four SiPMs on the amplitude of the
      light signal.
    }
    \label{fig:SiPM_RelResol}
\end{figure}

A series of resolution measurements were conducted on a photodetector
consisting of a single SiPM and a photodetector consisting of four
SiPMs, connected using the hybrid circuit described above. To measure
the resolution as a function of the photodetector signal amplitude, an
LED generator was used, with the pulse amplitude varied over the
entire range of the 12-bit DAC, i.e., from 1 to 4095.
Fig.~\ref{fig:SiPM_RelResol} shows the dependences of the relative
amplitude resolution $\sigma/A$ on the amplitude of the measured
signal for photodetectors made of one and four SiPMs.
The 1:2 resolution ratio for these two photodetectors is observed
across the entire amplitude range, confirming that resolution depends
on the square root of the photodetector area ratio or the number of
SiPMs installed on the photodetectors. This important conclusion about
the dependence of resolution on the number of SiPMs in a photodetector
can be used to extrapolate to larger numbers of SiPMs. For example, a
photodetector consisting of nine SiPMs should have a threefold
improvement in amplitude resolution compared to a photodetector with a
single SiPM, and so on.

A photodetector consisting of four SiPMs was used for spectral
analysis of amplitudes from the prototype counter of the GNT
spectrometer of gamma-nuclear transitions of the Hyperon+
experiment~\cite{Gorin:2024xsl} 
based on a NaI(Tl) crystal, which was irradiated with a $^{22}$Na
$\gamma$-source during the spectrometer calibration. The $^{22}$Na
emission spectrum contains two lines with energies of 511 and
1274.5~keV. The NaI(Tl) crystal is a cylinder with dimensions
$D/L=63/63$~mm.
The photodetector was mounted on the transparent end of the encapsulated crystal. Optical contact between the photodetector and the crystal was ensured by Saint Gobain optical grease.
The opposite end and the side of the crystal had matte white
reflectors to isotropically reflect the scintillation light emerging
from the crystal.
Since the total area of the four SiPMs, equal to $155~\mbox{мм}^2$, is 20 times
smaller than the area of the $3117~\mbox{мм}^2$ crystal end cap on which the
photodetector is mounted, 95\% of the scintillation light emitted by
the crystal escape from the photodetector. To improve light collection, a
reflector made of white matte Tyvek was mounted on the end cap of the
crystal. This reflector is a 63 mm diameter circle with a square
cutout for the four SiPM photodetector. This reflector improves light
collection by 1.5 times. 

\begin{figure}
  \centering
    \includegraphics[width=0.9\linewidth]{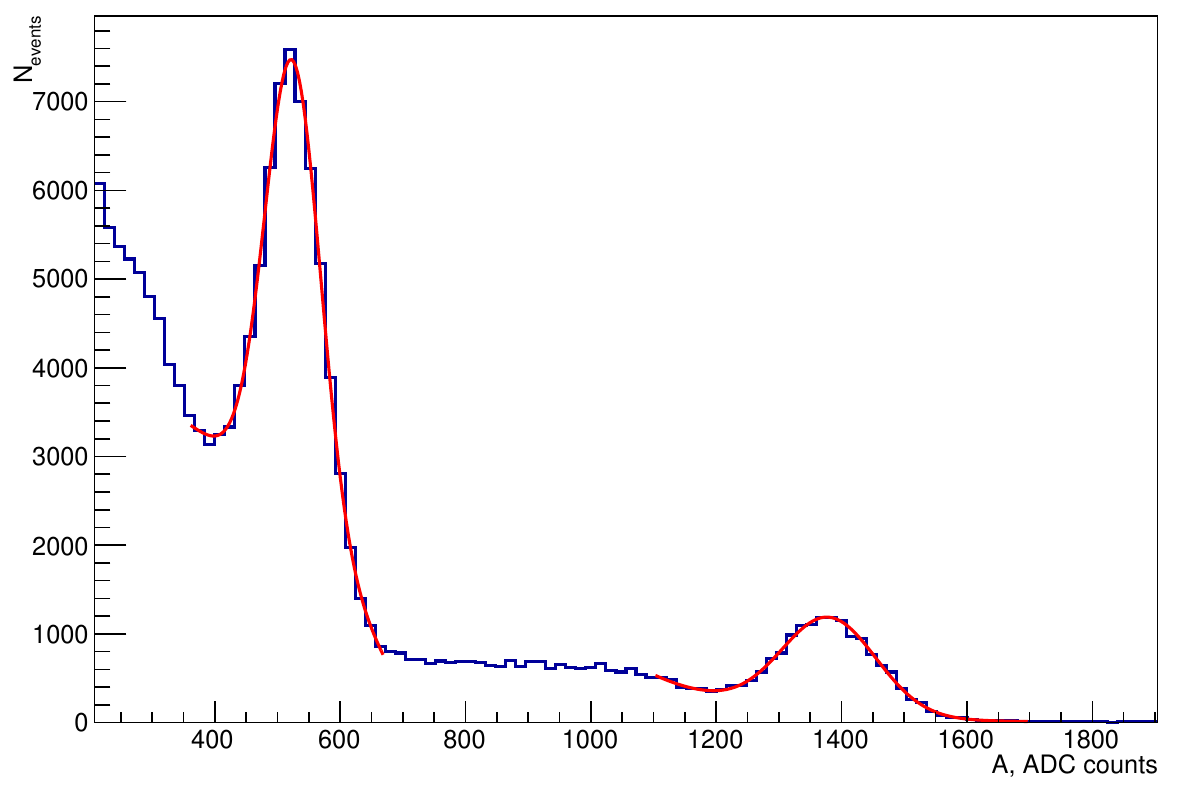}
    \caption{Fig. 6. Amplitude spectrum from a counter based on a
      NaI(Tl) crystal with a photodetector in the form of a $2\times2$
      SiPM matrix, irradiated with a $\gamma$-source $^{22}$Na.
    }
    \label{fig:GNT-SiPM}
\end{figure}

The counter prototype was integrated into the GNT detector
infrastructure. During detector calibration, signals from the counter
were fed for digitization to a 12-bit charge-sensitive ADC module
EM-6~\cite{Soldatov:2019cuc}, implemented in the Euro-MISS
standard~\cite{Bukreeva:2014wsa}. Due to the small area of the
photodetector, the number of photoelectrons recorded by the counter
was insufficient for measurements in the specified energy range, i.e.,
up to 5 MeV.
Therefore, the signal from the photodetector was fed to the ADC
through a low-noise amplifier with a gain of 30. The resulting
spectrum of digitized amplitudes is shown in Fig.~\ref{fig:GNT-SiPM},
where two peaks are clearly visible, at 524 and 1376 ADC counts,
corresponding to the $\gamma$ lines of 511 and 1274.5~keV of the
$^{22}$Na source. Fitting the peaks with Gaussian functions with a
polynomial background gives a relative energy resolution equal to
$$
\left. \frac{\sigma}{A}\right|_{511~\mbox{\scriptsize{keV}}} = 8.8\%, \quad
\left. \frac{\sigma}{A}\right|_{1274~\mbox{\scriptsize{keV}}} = 5.3\%,
$$
which is a bit worse than the resolution for the same counter, but
with a FEU-184TD as a photodetector, see~\cite{Gorin:2024xsl}. At the
same time, taking into account the effect of improving the counter
resolution with an increase in the photodetector area, one can expect,
for example, that for a photodetector in the form of a $4\times4$~SiPM
matrix, the resolution will be close to the resolution obtained in the
case of the traditional use of spectrometric PMTs as photodetectors.

\section{Conclusion}

Using the EQR15 11-6060D-S SiPM as an
example, this study demonstrates that large-area SiPMs with a 15-$\mu$m
pixel size can be used to detect light pulses over a wide range of
light fluxes with sufficient amplitude resolution and
linearity. Combining several SiPMs into a single photodetector using a
hybrid schematics can increase the light collection efficiency when
detecting light fluxes over a large area.
Moreover, the relative amplitude resolution of a SiPM array improves
with increasing total active area $S_A$ of the photodetector or the
number of combined SiPMs as $\sim 1/\sqrt{S_A}$. Fine-mesh SiPMs are
effective for detectors designed to measure signal amplitude over a
wide dynamic energy range, for example, in electromagnetic
calorimeters. SiPMs can also be used in detectors for gamma-nuclear
spectroscopy based on scintillating crystals with a high light
yield. Such photodetectors allow the measurement of signal amplitudes
varying by more than four orders of magnitude.

Using the prototype of the gamma-nuclear transition detector of the
Hyperon+ experiment based on a NaI(Tl) crystal, it was demonstrated
that a photodetector in the form of a $2\times2$ matrix of SiPM EQR15
S14160-6010PS, combined in a hybrid scheme, provides a resolution of
the gamma lines of the radioactive source $^{22}$Na 511~keV and 1274.5~keV
at a level of 8.8\% and 5.3\%, respectively. Further improvement of
the GNT detector resolution is possible by increasing the matrix of
SiPMs combined in a single photodetector.

\bibliographystyle{amsrusunsrt}
\bibliography{SiPM4spectroscopy.bib}

\end{document}